\documentclass[10pt]{article}
\usepackage[utf8]{inputenc}                                                     
\usepackage[margin=0.5in]{geometry}                                             
\usepackage{amsmath}                                                            
\usepackage{amsfonts}                                                           
\usepackage{graphics}
\usepackage{cite}

\usepackage{graphicx}
\usepackage{wrapfig}
\usepackage{lipsum}
\usepackage{lmodern}

\usepackage{tikz}

\usetikzlibrary{decorations.pathmorphing}
\usetikzlibrary{arrows,shapes,positioning}

\usetikzlibrary{decorations.pathmorphing}
\tikzset{snake it/.style={decorate, decoration=snake},
    photon/.style={decorate, decoration={snake}, draw=red},
    electron/.style={draw=blue, postaction={decorate},
        decoration={markings,mark=at position .55 with {\arrow[draw=blue]{>}}}},
    gluon/.style={decorate, draw=magenta,
        decoration={coil,amplitude=4pt, segment length=5pt}} 
}

\tikzset{
photon/.style={decorate, decoration={snake}, draw=red},
particle/.style={draw=blue, postaction={decorate},
    decoration={markings,mark=at position .5 with {\arrow[draw=blue]{>}}}},
antiparticle/.style={draw=blue, postaction={decorate},
    decoration={markings,mark=at position .5 with {\arrow[draw=blue]{<}}}},
gluon/.style={decorate, draw=black,
    decoration={coil,amplitude=4pt, segment length=5pt}}
 }

\long\def\ca#1\cb{} %Use for commenting out: \ca...\cb

\newcommand{\becs}{\begin{cases}}
\newcommand{\bem}{\begin{matrix}}

\newcommand{\bsk}{\bigskip }

\newcommand{\encs}{\end{cases}}
\newcommand{\enm}{\end{matrix}}

\newcommand{\dbar}{d\hspace*{-0.08em}\bar{}\hspace*{0.1em}}

 \def\outl#1{}     
\begin{document}
\begin{center}
{\bf Kinetics of Vesicle Growth}\bsk\\
\normalsize Bijit Singha \footnote{bsingha@andrew.cmu.edu} \\
\textit{Department of Physics, Carnegie Mellon University, Pittsburgh, PA 15213}\\
Date: \today
\vspace{.1cm}
\end{center}

 \begin{abstract}
A mechanism is proposed for the growth of vesicles dispersed in a liquid solvent and a size distribution function is obtained for the vesicles, both from the first principles calculations. This distribution function is shown to be positively skewed and evolving in time obeying a Fokker-Planck type equation. The critical size of the spherical vesicles is shown to grow in time with an exponent of 0.25. A constant is suggested that characterizes how easily a vesicle can absorb amphiphiles into its periphery.
\end{abstract}

\tableofcontents

\section{Introduction}
\label{sct1} 
Amphiphilic molecules can aggregate in a solvent to form different interesting structures. The structure of our interest is called vesicles, in which the amphiphilic bilayer can form hollow sphere. The medium inside such a vesicle can have the same or different concentration than the solvent outside but, for our purpose, we will consider same concentration of the solvent both inside and outside the vesicle. The natural assemblance of the amphiphilic bi-molecules to form such symmetric structures, due to its simplicity and homogeneity in composition, can work as a potential tool to understand the dynamics at the mesoscopic scale. Consequently, these systems have been a subject of wide theoretical studies.\\

\noindent
Vesicles were observed for the first time more than fifty years ago\cite{VesicleDiscovery1965} and since then, they have been used to improve our understanding of a number of cellular processes\cite{PantazatosDP1999, CansAS2003,LeiG2008, Lapinski2007}, and also in various industrial and commercial applications\cite{Fernandez2005}. Along with experimental progress, there have been many important theoretical development in understanding the formation and growth of this biological system. Helfrich, in his famous work on the lipid bilyer elasticity\cite{Helfrich1973}, derived the curvature contribution to the free energy and worked on ellipsoidal deformation of the spherical vesicular structure by magnetic fields. Morse and Milner, in their work \cite{Morse1994}, added the finite size contributions to the free energy derived by Helfrich. They showed that the free energy depends on the vesicle size logarithmically. There were several important works before this as well, arguing the existence of the $c\ln N$ contribution to the free energy\cite{Helfrich1986,Huse1988}, with $c$ being a constant and is less than zero and $N$ being the number of bi-molecules in the vesicle. Morse and Milner found that $c$ is positive with a value of $7/6$ (we will show in a later section how the signature of $c$ will be constrained from the drift of the vesicle size distribution). Several models have been proposed in literature on the positively-skewed size distribution of the vesicles in a solvent\cite{Morse1994,Guida2010} as well.\\

\noindent
This work is motivated by a recent observation that the mode of the vesicle size distribution grows in time with a scaling exponent of $1/4$ for some vesicles\cite{Paulaitis2018}. This was found by assuming that the characteristic scale associated with this process has a power-law dependence on time, $l(t) \sim t^\xi$, and then $\xi$ was fitted to experimental data. In this work, we find out, from first principles calculations that, for an ideal spherical vesicle, the scaling exponent of the exosome size growth is indeed $1/4$ for the critical radius. This helps us to develop a formalism to understand the equilibrium size distribution of vesicles at a given time and its time-evolution of such a system that contains vesicles in aggregate of different sizes. We obtain a Fokker-Planck (FP) equation that will dictate the dynamics of the vesicle-growth process. FP equation has been previously used in \cite{Zapata2009} to understand the formation of spherical vesicles but we will focus primarily on vesicle-growth in this work. To characterize how easily an amphiphilic bi-molecule can assimilate into the periphery of a given vesicle, we also propose a constant in this work. This constant will be determined by the detailed chemical and structural properties of any vesicle.\\

\noindent
We assume in this work that the dissolution or the growth of the vesicles proceeds through the emission or assimilation of a single amphiphile bi-molecule from the periphery of the vesicles. We have also used the principles of detailed balancing which will hold only when the system is in thermal equilibrium states at all time. Both the assumptions are justified because the vesicle-growth, as found from the experiments, is a sufficietly slow process. Moreover, several interesting properties of the system have direct correspondences to the specific features of the thermodynamic functions, such as free energy. This helps us in two ways: (i) we can come up with the constraints that these functions should satisfy by simply analyzing their consequences to a real system and comparing with the observations, (ii) finding the effect of such functions on the size and time-evolution of the system gives us an intuitive understanding of the corresponding system. This also allows to specify the constants characterizing such processes ($e.g.$ coefficiient of assimilation) from the constraints imposed by observations about the vesicle sizes and their time-evolution.\\

\noindent
The structure of this paper is as follows. In Section~\ref{sct2}, we will find out that the chemical potential corresponding to the vesicle growth processes is effectively a constant for large size of vesicles. In Section~\ref{sct3}, we will assume a size distribution function for the vesicles in the growth phase and will derive a Fokker-Planck equation using principles of detailed balancing. This will give us an FP equation for the time evolution of the vesicle size distribution in general. In Section~\ref{sct4}, we will use this equation for the special case of spherical vesicles to derive the scaling exponent $0.25$ of time for the growth process. We will discuss our findings in Section~\ref{sct5}.

\section{Thermodynamics of vesicle formation and growth}
\label{sct2} 

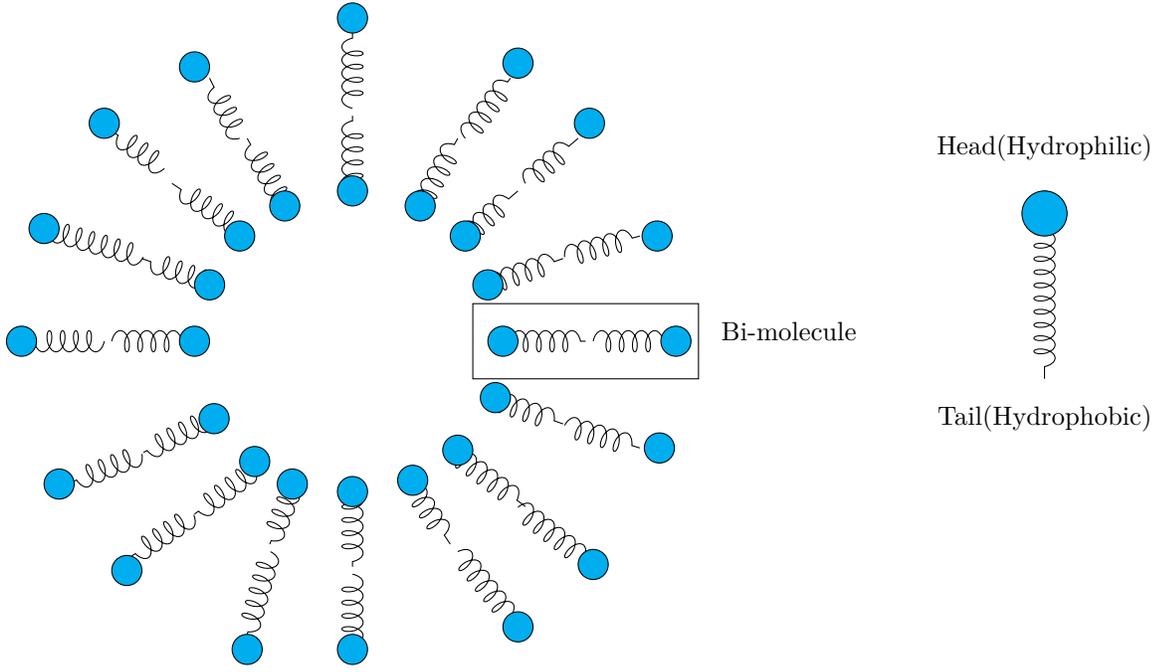
\begin{figure}
\begin{tikzpicture}
\coordinate (O) at (0,0,0);
\coordinate (A) at (2,0,0);
\coordinate (B) at (0,1,0);
\coordinate (C) at (2,1,0);

% \draw (2.4,0.5) circle (2cm);
% \draw (2.4, 0.5) circle (4.2cm);

\node[label=below:Bi-molecule] at (8.2,1){};

\draw (7,0) rectangle (4,1);

 \draw [gluon] (4.5,0.5) -- (5.5,0.5);  \draw [fill = cyan] (4.4,0.5) circle (0.2cm);
 \draw [gluon] (5.6,0.5) -- (6.6, 0.5); \draw [fill = cyan] (6.7, 0.5) circle (0.2cm);
  \draw [gluon] (-0.8,0.5) -- (0.2, 0.5); \draw [fill = cyan] (0.3, 0.5) circle (0.2cm);
  \draw [gluon] (-0.9,0.5) -- (-1.9, 0.5); \draw [fill = cyan] (-2.0, 0.5) circle (0.2cm);
  
  \draw [gluon] (2.4,2.5) -- (2.4, 3.5);  \draw [fill = cyan] (2.4,2.5) circle (0.2cm);
  \draw [gluon] (2.4,3.6) -- (2.4, 4.6);    \draw [fill = cyan] (2.4,4.8) circle (0.2cm);
  \draw [gluon] (2.4, -1.5) -- (2.4, -2.5);   \draw [fill = cyan] (2.4,-1.5) circle (0.2cm);
  \draw [gluon] (2.4, -2.6) -- (2.4, -3.6);  \draw [fill = cyan] (2.4,-3.6) circle (0.2cm);
  
  \draw [gluon] (4.2,1.25) -- (5.2, 1.6);  \draw [fill = cyan] (4.2,1.25) circle (0.2cm);
  \draw [gluon] (5.22,1.62) -- (6.22,1.9);  \draw [fill = cyan] (6.45,1.9) circle (0.2cm);
   \draw [gluon] (0.6,1.25) -- (-0.4, 1.6);   \draw [fill = cyan] (0.5,1.25) circle (0.2cm);
   \draw [gluon] (-0.38,1.62) -- (-1.7,2);  \draw [fill = cyan] (-1.7,2) circle (0.2cm);
   
   \draw [gluon] (-0.4,-1) -- (-1.4, -1.4);  \draw [fill = cyan] (-1.5,-1.4) circle (0.2cm);
    \draw [gluon] (0.5,-0.53) -- (-0.38,-0.98);  \draw [fill = cyan] (0.56,-0.53) circle (0.2cm);
   \draw [gluon] (4.2,-0.25) -- (5.2,-0.6);  \draw [fill = cyan] (4.3,-0.25) circle (0.2cm);
   \draw [gluon] (5.22, -0.62) -- (6.22, -0.92); \draw [fill = cyan] (6.48,-0.92) circle (0.2cm);
   \draw [gluon] (3.7, -0.95) -- (4.7, -1.7);  \draw [fill = cyan] (3.8, -0.95) circle (0.2cm);
   \draw [gluon] (4.6,-1.71) -- (5.6,-2.49); \draw [fill = cyan] (5.6,-2.47) circle (0.2cm);
   \draw [gluon] (3.2,-1.45) -- (3.7, -2.2); \draw [fill = cyan] (3.2,-1.35) circle (0.2cm);
    \draw [gluon] (3.8, -2.3) -- (4.6,-3.3);  \draw [fill = cyan] (4.6,-3.3) circle (0.2cm);
   \draw [gluon] (1.6,-1.45) -- (1.3,-2.2); \draw [fill = cyan] (1.6,-1.4) circle (0.2cm);
    \draw [gluon] (1.31,-2.3) -- (1.0,-3.4);  \draw [fill = cyan] (1,-3.6) circle (0.2cm);
   \draw [gluon] (1.1,-1.2) -- (0.3,-1.8);  \draw [fill = cyan] (1.1,-1.1) circle (0.2cm);
   \draw [gluon] (0.29,-1.81) -- (-0.60, -2.4);  \draw [fill = cyan] (-0.60,-2.55) circle (0.2cm);
   \draw [gluon] (0.9,2) -- (0.0,2.6);  \draw [fill = cyan] (0.9,1.9) circle (0.2cm);
   \draw [gluon] (-0.1, 2.8) -- (-0.8,3.3);  \draw [fill = cyan] (-0.9,3.4) circle (0.2cm);
   \draw [gluon] (1.5,2.3) -- (1,3.2);  \draw [fill = cyan] (1.5,2.3) circle (0.2cm);
   \draw [gluon] (0.9,3.2) -- (0.5, 4.0);  \draw [fill = cyan] (0.3,4.15) circle (0.2cm);
  
   \draw [gluon] (3.3,2.3) -- (3.8,3.2);  \draw [fill = cyan] (3.3,2.3) circle (0.2cm);
    \draw [gluon] (3.9,3.2) -- (4.5,4.0);  \draw [fill = cyan] (4.6,4.2) circle (0.2cm);
   \draw [gluon] (3.9,1.9) -- (4.6,2.5);  \draw [fill = cyan] (3.9,1.9) circle (0.2cm);
   \draw [gluon] (4.7,2.6) -- (5.4,3.2);  \draw [fill = cyan] (5.55,3.4) circle (0.2cm);
   
\node[label=below:Head(Hydrophilic)] at (11.6,3.5){}; 
\node[label=below:Tail(Hydrophobic)] at (11.6,-0.1){}; 
   
   \draw [gluon] (11.6,2.5) -- (11.6,0); \draw [fill = cyan] (11.6,2.2) circle(0.3cm);

\end{tikzpicture}

\caption{2-dimensional projection of a vesicle(on the left) with amphiphilic bi-layer. If this vesicle has $i$ number of amphiphilic bi-molecules, it will belong to aggregate-$i$. Each amphiphile(on the right) has a hydrophobic tail and a spherical hydrophilic head of radius $a$.}
\end{figure}

When a large number of amphiphile bi-molecules are dissolved, vesicles can be formed in the solvent. In this process, entropy of the solute amphiphiles increases along with an increase in their interaction free energy. The former is due to the increase in the count of all possible microstates of the system, while the second effect is attributed to the fact that the amphiphiles like to be surrounded by similar bi-molecules than the solvent molecules. Structure formation by these amphiphiles becomes possible when the interaction energy dominates the energy associated with the increase in entropy. In what follows in this section, we will show that when such vesicles are formed, the chemical potential of the vesicles is the same for the sufficiently large size.\\
Let's imagine that vesicles in aggregate of different sizes $\{i\}$ are present in the solvent, with $i$ being the number of amphiphile bi-molecules in a given vesicle, and let's say there are $N_i$ number of vesicles in aggregate-$i$. From the first law of thermodynamics, we can write
\begin{eqnarray}
\label{1stLaw}
\dbar Q = dU + PdV,
\end{eqnarray}
where $\dbar Q$ is the heat gained by the system, $U$ is the internal energy, $P$ is the pressure and $V$ is the volume. Because the system is not thermally isolated in general, second law gives us
\begin{eqnarray}
dU + P dV \leq TdS,
\end{eqnarray}
where $S$ is the entropy of the system. Because we have different sizes of vesicles, the energy density of the system will depend on the relative amounts of the aggregates and thus, we have to add one more term to the above expression to write
\begin{eqnarray}
\label{InEq}
d U + PdV - \sum_i \mu_i dN_i \leq T dS,
\end{eqnarray}
Here $\mu_i$ denotes the chemical potential of a vesicle that belongs to the aggregate-$i$. The change in Gibbs Free Energy $dG = dU + pdV - TdS$ should be less than zero for any spontaneous process (and is zero for processes at equilibrium). Combining this with Eq.~(\ref{InEq}), we get
\begin{eqnarray}
\sum_i \mu_i dN_i \leq 0.
\end{eqnarray}
Because vesicle growth is a very slow process, we assume that this is in equilibrium:
\begin{eqnarray}
\sum_i \mu_i dN_i = 0.
\end{eqnarray}

%----------------------BEGIN COMMENT----------------------------

%For any spontaneous process or a process in equilibrium at constant temperature and pressure, Gibbs Free Energy obeys the following relation:
%\begin{eqnarray}
%d G \leq 0,
%\end{eqnarray}
%and thus, from Eq.~(\ref{InEq}), we get
%\begin{eqnarray}
%V dP - S dT + \sum_i \mu_i dN_i \leq 0.
%\end{eqnarray}
%Now, for the system to be in equilibrium, this free energy should be minimum at constant temperature and pressure. This tells us
%\begin{eqnarray}
%\sum_i \mu_i dN_i = 0.
%\end{eqnarray}
%----------------------END COMMENT-------------------------------

Now, if some vesicles of aggregate-$i$ absorb the same number of vesicles of aggregate-$j$ to form vesicles of aggregate-$(i+j)$, then $-dN_i = -dN_j = +dN_{i+j}$. Using this condition in the above equation, we get
\begin{eqnarray}
\label{AdditionOfPotential}
\mu_i + \mu_j = \mu_{i+j}.
\end{eqnarray}
From the above expression, we can write
\begin{eqnarray}
\mu_i + \mu_{j + \delta n} = \mu_{i + j + \delta n}.
\end{eqnarray}
Taylor-expanding the second term on the left hand side and the term on the right hand side and ignoring the higher order terms $\mathcal{O}(\delta n^2)$ for $\delta n \ll i, j$, we get
\begin{eqnarray}
\mu_i + \mu_j + \delta n \frac{d \mu_j}{d n} = \mu_{i+j} + \delta n\frac{d \mu_{i+j}}{d n}
\end{eqnarray}
Using Eq.~(\ref{AdditionOfPotential}) again gives us
\begin{eqnarray}
\label{MuNConstancy}
\mu_n = \rm{constant}
\end{eqnarray}
for sufficiently large $n$. This is a very important relation and will determine the equilibrium size distribution of the vesicles. This tells us that the large-sized vesicles will be in equibrium and will not like to exchange amphiphiles, but they can grow in size by absorbing bi-molecules from small vesicles. This will lead to a specific scaling exponent in time-dependence of the distribution function, as will be shown in the next section.

\section{Size distribution function of the vesicles}
\label{sct3}
\noindent
We assume a distribution function for a vesicle of volume $V$, containing $n$ bi-molecules at time $t$. This is similar to a probability density function at a given time if normalized to unity:
\begin{eqnarray}
\label{SizeDist}
f(n,V,t) = \psi(n,t) \delta(V - V(n)),
\end{eqnarray}
where the delta function is basically implying that the volume of the vesicle is determined by the number of bi-molecules on its surface. The size-dependence and the time evolution of this function will be dictated by thermodynamics. The size of the vesicle changes by aggregation or dissociation of amphiphile molecules and can be expressed by a kinetic equation
\begin{eqnarray}
\frac{\partial f(n,t)}{\partial t}
=
\alpha^{(+)}_{n-1,n} f(n-1,t) - \alpha^{(-)}_{n,n-1} f(n,t) + \alpha^{(-)}_{n+1,n} f(n+1,t) - \alpha^{(+)}_{n,n+1} f(n,t),
\end{eqnarray}
where $\alpha^{(+)}_{i,i+1}$ is the average number of events per unit time that one bi-molecule is absorbed by the vesicle to increase the total number from $i$ to $i+1$, $\alpha^{(-)}_{i,i-1}$ is the average number of events per unit time that one bi-molecule is released from the vesicle to decrease the total number from $i$ to $i -1$. Defining the flux $J_n$ as
\begin{eqnarray}
J_n = \alpha^{(+)}_{n,n+1}f(n,t) - \alpha^{(-)}_{n+1,n}f(n,t),
\end{eqnarray}
we can write
\begin{eqnarray}
\frac{\partial f}{\partial t} &=&
- \left(
J_n - J_{n-1}
\right).
\end{eqnarray}
The absorption and emission coefficients $\alpha^{(+)}$ and $\alpha^{(-)}$ can be determined from the kinetics of the corresponding process. For a vesicle of critical radius, aggregation and dissociation cancels out each other exactly and assuming the following equilibrium distribution of the vesicle
\begin{eqnarray}
f^{(eq)}(n)
&\sim&
 \exp \left[
- \frac{R_{rev}(n)}{k_B T}
\right],
\end{eqnarray}
where $R_{rev}(n)$ is the work of formation of a vesicle consisting of bi-molecules done in a reversible process. For constant temperature and pressure, this is basically the Gibbs free energy:
\begin{eqnarray}
R_{rev}(n) = \Delta G(n).
\end{eqnarray}
From this point onwards, we will call $\Delta G$ as free energy. Now, we use the principle of detailed balancing to write
\begin{eqnarray}
\frac{\alpha^{(+)}_{n,n+1}}{\alpha^{(-)}_{n+1,n}}
&=&
\frac{f^{(eq)}(n+1)}{f^{(eq)}(n)} \nonumber\\
&=&
\exp \left[
- \left(
\frac{\Delta G(n+1) - \Delta G(n)}{k_B T}
\right)
\right].
\end{eqnarray}
Using these will eventually lead us to a Fokker-Planck equation that can describe time-evolution of a vesicle during the nucleation and growth processes:
\begin{eqnarray}
\label{FokkerPlanck}
\frac{\partial f(n,t)}{\partial t}
= - \frac{\partial J}{\partial n},
\end{eqnarray}
where
\begin{eqnarray}
\label{DefJ}
J(n,t) = - \alpha ^{(+)}_{n,n+1} \left[
\frac{\partial f(n,t)}{\partial n}
+ \frac{f(n,t)}{k_B T} \frac{\partial \Delta G(n)}{\partial n}
\right].
\end{eqnarray}
Similar expressions are used for a lot of other thermodynamic systems, as shown in \cite{SlezovBook}. The first term in Eq.~(\ref{DefJ}) is the diffusion flow rate and the second term corresponds to thermodynamic flow rate or the drift flow rate. The first term in Eq.~(\ref{DefJ}) describes the size distribution cloud spreading diffusively while the second term denotes the peak of the same cloud drifting towards a different $n$-value. In the absence of the second term, this becomes a zero-drift equation with constant diffusion and the peak of the size distribution simply spread out but doesn't move. Thus, we can see from Eq.~(\ref{DefJ}) that it is the size-dependence of the free energy which is pushing the peak of the distribution function to the right and thus dictates the critical growth rate of the vesicles.\\

\noindent
Let's recall briefly a few ideas about stochastic processes such as the one we are dealing with in this work. We can consider a stochastic differential equation of the form
\begin{eqnarray}
\label{DefEto}
d\textbf{n} = \boldsymbol{\mu}(\textbf{n},t) dt + \boldsymbol{\sigma}(\textbf{n},t) d\textbf{W}_t
\end{eqnarray}
where $\boldsymbol{\mu}$ is a $K$-dimensional vector, $\boldsymbol{\sigma}$ is an $K \times K$ matrix and $d\textbf{W}_t$ is a $M$-dimensional vector that denotes the infinitesimal increment in the Wiener process with the property that
\begin{eqnarray}
\langle \textbf{W}_t \rangle &=& 0,\\
\langle \textbf{W}_t \textbf{W}_{t'} \rangle &=& {\rm{min}}(t,t').
\end{eqnarray}
Any process corresponding to Eq.~(\ref{DefEto}) is called It\^o process. Corresponding to this process, there exists a Fokker-Planck equation satisfied by the conditional probability density $P(\textbf{n},t\vert \textbf{n}_0, t_0)$ of the random variable $N_t$:
\begin{eqnarray}
\label{FPfromIto}
\frac{\partial P(\textbf{n},t)}{\partial t} &=&
- \sum_{i = 1}^K \frac{\partial}{\partial n_i} \left[
\mu_i(\textbf{n},t)P(\textbf{n},t)
\right]
+ \sum_{i = 1}^K \frac{\partial^2}{\partial n_i \partial n_j} \left[
D_{ij}(\textbf{n},t) P(\textbf{n},t)
\right]
\end{eqnarray}
%This can be proven from the Ito calculus. 
In Eq.~(\ref{FPfromIto}), $\boldsymbol{\mu}$ is called the drift vector and $\textbf{D} = \frac{1}{2}\boldsymbol{\sigma} \boldsymbol{\sigma}^T$ is called diffusion tensor. We have used tensor indices in Eq.~(\ref{FPfromIto}) to show the dimensionality of each term explicitly. Defining the diffusion current using Fick's law
\begin{eqnarray}
J^i_{\rm{diff}}(\textbf{n},t) = - D^{ij} \frac{\partial P(\textbf{n},t)}{\partial n^i}
\end{eqnarray}
and the drift current 
\begin{eqnarray}
\label{DefDrift}
J_{\rm{drift}}^i(\textbf{n},t) = \mu^i(\textbf{n}) P(\textbf{n},t),
\end{eqnarray}
we can write the Fokker-Planck equation, Eq.~({\ref{FPfromIto}}), as
\begin{eqnarray}
\label{FPinDriftAndDiffusion}
\frac{\partial P(\textbf{n},t)}{\partial t} &=&
- \frac{\partial}{\partial n^i} \left[
J^i_{\rm{drift}}(\textbf{n},t) + J^i_{\rm{diff}}(\textbf{n},t)
\right]
\end{eqnarray}
Comparing Eq.~(\ref{DefDrift}) and (\ref{FPinDriftAndDiffusion}) with Eq.~(\ref{FokkerPlanck}) and Eq.~(\ref{DefJ}), we get that the drift current
\begin{eqnarray}
J_{\rm{drift}}(n,t) = \frac{\alpha_{n,n+1}}{k_B T} \frac{\partial \Delta G}{\partial n} f(n,t)
\end{eqnarray}
and the drift vector
\begin{eqnarray}
\label{DriftinFP}
\mu(n,t) = \frac{\alpha_{n,n+1}}{k_B T} \frac{\partial \Delta G}{\partial n}.
\end{eqnarray}
Notice that this framework can be applied generally for any vesicle systems. In the next section, we are going to assume spherical vesicles and derive a more specific equation governing the time-evolution of such a system. 

\section{Time-evolution of the size distribution}
\label{sct4}
\noindent
We can see from Eq.~(\ref{DefJ}) that we need the free energy of the vesicle in order to find out the size distribution and the dynamic scaling of the vesicle. Free energy is different for different structures but we consider here, for simplicity, an aggregate of spherical vesicles. We use the same form of the free energy for a spherical vesicle as obtained in \cite{Morse1994}:
\begin{eqnarray}
\label{FreeEnergy}
\Delta G(n) = 8 \pi (\kappa + \frac{1}{2}\bar{\kappa}) + k_B T ({\rm{const.}}) - c k_BT \ln n~.
\end{eqnarray}
In the above expression, $\kappa$ and $\bar{\kappa}$ are the vesicle's bending rigidity and gaussian rigidity, $T$ is the temperature of the system, $n$ is the number of bi-molecules in a given vesicle, $k_B$ is the Boltzmann constant and $c$ is another constant. Because the first term in the right hand side is scale-independent, the phase behavior is dictated only by size-dependent contribution coming from the last term in that expression. Notice that, the logarithmic dependence of that term tells us that the chemical potential is effectively a constant near the critical radius of the vesicle, as expected from Eq.~(\ref{MuNConstancy}).  Using Eq.~(\ref{FokkerPlanck}) and (\ref{DefJ}), and assuming that the diffusivity remains constant during the growth process ($\alpha_{n,n+1} = D_0$ for sufficiently large $n$), we get
\begin{equation}
\label{FP}
\frac{\partial \psi}{\partial t}
=
D_0 \frac{\partial}{\partial n} \left[
\frac{\partial \psi}{\partial n} + \frac{\psi}{k_B T} \frac{\partial \Delta G}{\partial n}
\right]~.
\end{equation}
Comparing this expression with Eq.~(\ref{FPfromIto}), (\ref{DefDrift}), (\ref{DriftinFP}) with the above equation, we get the drift speed of the size distribution function assumed in Eq.~(\ref{SizeDist}):
\begin{eqnarray}
u(n,t) =
\frac{\partial n_M}{\partial t}
&=&
\frac{c D_0}{ n_M}~.
\end{eqnarray}
Here $n_M$ denotes the mode size at a given time. Thus,
\begin{eqnarray}
n_M(t) = \sqrt{2cD_0}~ t^{1/2} \sim t^{1/2}~.
\end{eqnarray} 
 
\noindent Comparing the above expression with the experimental result, we can get the diffusivity constant corresponding to this process. We can express the same differential equation in terms of the radius of the vesicle using 
\begin{equation}
\label{NRRelation}
n = \frac{4 \pi R^2}{\pi a^2} = \frac{4R^2}{a^2},
\end{equation}
where $R$ is the radius of the vesicle and can be measured experimentally, and $a$ is the radius of the cross-section of the hydrophilic head of the vesicle. This gives the the dependence of the critical radius on time:
\begin{eqnarray}
R_c = \frac{a}{2} \left( \frac{7 D_0}{3} t  \right)^{1/4} \sim~ t^{1/4}~.
\end{eqnarray}
The scaling exponent of time in the above equation is $0.25$.
It is worth mentioning here that, in \cite{Paulaitis2018}, the authors found a similar result by assuming a power-law scaling of the mean diameter and fitting the exponent to their experimental data.
Also, our observable is the size of the vesicles and it will be more helpful to express Eq.~(\ref{FP}) in terms of the radius of the vesicles.
\begin{equation}
d n = \frac{8R dR}{a^2},
\end{equation}
and thus, from Eq.~(\ref{FP}),
\begin{eqnarray}
\frac{\partial \psi}{\partial t} &=&
\frac{D_0a^4}{64 R^2}\left[
\frac{\partial^2 \psi}{\partial R^2} - \left(\frac{1+2c}{R}\right)\frac{\partial \psi}{\partial R} + \frac{4c\psi}{R^2}
\right]
.
\end{eqnarray}

\section{Conclusion}
\label{sct5}
We have given a framework under which we can calculate the time evolution of size distribution of the vesicles, and showed that the critical size of sphericle vesicles will grow with time with a scaling exponent of $0.25$. This is derived in a straightforward way, first by assuming that the free energy is only logarithmically (\textit{i.e.} slowly) varying with the number of bi-molecules, thus the chemical potential being a constant effectively for sufficiently large size of vesicles, and then finding out how the size distribution will evolve by writing down the Fokker-Planck equation corresponding to the process of bi-molecule exchange between different aggregates. Furthermore, we have proposed a constant to characterize the ease of diffusivity of an amphiphile bi-molecule into the periphery of the vesicle. The chemistry and the detailed structural properties will be encoded in this constant while the growth process will be dictated by thermodynamics. This means any deviation from the ideal value of this constant (\textit{i.e.} value obtained from this work for a spherical vesicle with same solvent with similar concentration inside and outside) can give us information about the physical or chemical interactions or any other form of perturbations playing a role in the system. All these behooves further investigation with more complicated vesicle systems. The scaling exponent will vary similarly for non-ideal structures but we can use the same underlying theory mentioned in this article to comment on the compositions and geometry of the corresponding vesicles, scope for future work.

\section*{Acknowledgements}
The author is thankful to Saigopalakrishna Yerneni and Sushil Lathwal for all the valuable discussions about vesicles and some other interesting biological systems, and to Prof. Changjin Huang and Prof. Markus Deserno for their valuable comments on this work. This work was supported by Carnegie Mellon University Department of Physics. 

%\appendix
%\section{Appendix. A}

\bibliographystyle{unsrt}

\end{document}